\documentclass[preprint, showpacs, preprintnumbers, amsmath,amssymb]{revtex4}

\usepackage{graphicx}
\usepackage{dcolumn}
\usepackage{tensor}
\usepackage{fontenc}

\begin{document}

\title{ The Friedmann equations in deformed Ho\v{r}ava-Lifshitz gravity and Debye model }
\author{Bin Liu}
\author{Yun-Chuan Dai}
\author{Xian-Ru Hu}
\author{Jian-Bo Deng}
\email{dengjb@lzu.edu.cn} \affiliation{
Institute of Theoretical Physics, Lanzhou University \\
Lanzhou 730000, People's Republic of China}

\date{\today}

\begin{abstract}

Considering the results from Ho\v{r}ava-Lifshitz (HL) theory, 
a more precise relation between the number of bits and area in the holographic system is proposed. 
With this corrected relation and Debye model, 
two modified Friedmann equations are derived 
from the Hawking temperature and the Unruh temperature separately in entropic force. 
These equations could be better in describing the whole evolution of the universe.


\end{abstract}
\maketitle

\section{introduction}

There are a number of evidences indicating a deep relation between gravity and thermodynamics.
The concept of gravity as a fundamental interaction has been challenged since 1967 \cite{f1}.
The socalled induced gravity which was proposed by Sakharov suggests that  spacetime background 
emerges as a mean field approximation of underlying microscopic degrees of freedom \cite{f2}. 
After thermodynamic properties of black hole discovered in 1970s, 
the entropy of a black hole is 1/4 of the area of the event horizon of the black hole measured in Planck units \cite{f3}.
This area law of entropy was extended to cover more general systems and later became a general principle called holographic principle \cite{f4}.
The strongest evidence of the holographic principle was provided by the AdS/CFT correspondence 
which states that all information about a gravitational system in a spatial region is encoded in its boundary \cite{f5}.
From other aspect, a deep connection between gravitation and thermodynamics is revealed by 
the thermodynamic laws of black holes and de-Sitter space-time \cite{f6}. \\
\indent An especially intriguing step forward in this context was Jacobson's realization \cite{f7}, that the argument also goes in the other direction: you can start with thermodynamics and derive the Einstein equations from there.
This issue has subsequently been discussed a lot by many authors. 
Padmanabhan \cite{f8} first noticed that the gravitational field equation in a static spherically symmetric spacetime can be rewritten as a form of the ordinary first law of thermodynamics at a black hole horizon.
This observation was then extended to the cases of horizons in gravity \cite{f9}\cite{f10}\cite{f11}\cite{f12}\cite{f13}\cite{f14}.
Furthermore, in the apparent horizon of a Friedmann-Robertson-Walker (FRW) universe 
Friedmann equations of the FRW universe can be derived with any spatial curvature by applying 
the Clausius relation to apparent horizon \cite{f15}. There are a lot of papers discussing the relation 
between the first law of thermodynamics and the Friedmann equations of FRW universe in variety of
gravity theories. Further references can be found in \cite{f16}\cite{f18} etc. \\
\indent Recently, Verlinde \cite{f19} contributed a remarkable new idea that gravity can be expressed
as an entropic force caused by the information changes when a material body moves away from 
the holographic screen which implies that gravity is not fundamental. 
Similar observations also arised in Padmanabhan's investigation \cite{f20}. 
Following the idea of entropic force, some applications have been performed. 
The Friemdann equations and the modified Friedmann equations for Friedmann-Robertson-Walker universe 
in Einstein gravity \cite{f21}\cite{f22}, f(R) gravity \cite{f23}, deformed Ho\v{r}ava-Lifshitz gravity \cite{f24}, 
and braneworld scenario \cite{f25} were derived utilizing the holographic principle and the equipartition rule of energy. 
The Newtonian gravity in loop quantum gravity was given by entropic force in Ref. \cite{f26}. 
In Ref. \cite{f27} the Coulomb force was regarded as an entropic force.
Ref. \cite{f28} showed that the holographic dark energy can be derived from the entropic force formula.
 J. Makela pointed out in \cite{f29} that Verlinde's entropic force is actually the consequence of a specific microscopic model of spacetime. 
The similar ideas were also applied to the construction of holographic actions from black hole entropy \cite{f30}
while Ref. \cite{f31} showed that gravity has a quantum informational origin. In \cite{f32}, 
a modified entropic force in the Debye model was presented . Another important discovery was made by
Joakim Munkhammar more recently \cite{f44},
which indicates an deep connection between the quantum mechanical probability density distribution and entropy.
For other relevant works to entropic force, we refer to e.g. \cite{f33}\cite{f34}\cite{f35}\cite{f17}\cite{f45}\cite{f46}and references therein.\\
\indent In this paper, we consider entropic force in the Debye model (MEF model) to derivate the modified
Friedmann equation of FRW universe in the deformed HL gravity. The letter is organized as follows. 
First, we review the model of modified Friedmann equation from HL gravity and the MEF model where a modified total number of bits is proposed.
Second, with the idea of Debye model together with 
the modified HL gravity, we obtain our modified Friedmann equation from the Hawking temperature. 
Third, with the same models we reproduce another modified Friedmann equation from the Unruh temperature.
Finally, the paper ends with a brief summary.
In this paper, we set the units of $c=G=\hbar=k_B=1$.


\section{THE MODEL OF MODIFIED FRIEDMANN EQUATION FROM HL GRAVITY AND THE MEF MODEL}\label{SecB}

To make our narration clear, we first review the model of modified Friedmann equation from HL gravity and the MEF model. Then, we apply it to the deformed HL gravity.\\
\indent Recently, Ho\v{r}ava proposed a new gravity theory which motivated by Lifshitz theory in solid state physics \cite{f36}, 
at a Lifshitz point \cite{f37}\cite{f38}\cite{f39}. The theory is usually referred to as the Ho\v{r}ava-Lifshitz (HL) theory.
It has manifest 3-dimensional spatial general covariance and time reparametrization invariance. This is a non-relativistic 
renormalizable theory of gravity and recovers the four dimensional general covariance only in an infrared limit. Thus, it may 
be regarded as a UV complete candidate for general relativity. The black hole solutions in this gravity theory have been 
attracted much attention. Much work in this field has been made \cite{f40}.\\
\indent It is interesting that there exists a logarithmic term in the entropy/area relation $S=\frac{A}{4}+\frac{\pi}{\omega}\ln A$  for the deformed HL gravity \cite{f13}\cite{f41}\cite{f42}. The logarithmic term may be regarded as a unique feature of black holes in the HL gravity. Considering this entropy/area relation, the modified Friedmann equation from the first law of thermodynamics could be obtained in \cite{f24}. Their modified Friedmann equation:
\begin{equation}\label{eq1}
(H^2+\frac{K}{a^2})+\frac{1}{2\omega}(H^2+\frac{K}{a^2})^2=\frac{8 \pi \rho }{3}
\end{equation}
Here $\rho$ is the total energy density; $K$ is the spatial curvature of the universe; $H=\dot{a} / a $ is the Hubble parameter; $a$ is the scale factor. Eq (\ref{eq1}) has a solution
\begin{equation}\label{eq2}
H^2+\frac{K}{a^2}=\omega(-1+\sqrt{1+\frac{16\pi\rho}{3\omega}})
\end{equation}
If $\rho \gg \frac{3\omega}{16\pi} $, we can ignore 1 and -1 at the right hand side of (\ref{eq2}) and get
\begin{equation}\label{eq3}
H^2+\frac{K}{a^2}=\sqrt{\frac{16\pi\rho}{3}}
\end{equation}
If $\rho \ll \frac{3\omega}{16\pi} $, we then can expand the right hand side of (\ref{eq2}) and get
\begin{equation}\label{eq4}
H^2+\frac{K}{a^2}=\frac{8\pi\rho}{3}
\end{equation}
\indent It has been shown that their modified Friedmann equation will back to the standard Friedmann equation at small $\rho$ approximation, which is applicable to  late Universe with low energy density. In next section, we will derivate another modified Friedmann equation using this model.\\
\indent It is important to note that the entropy-area relation plays a crucial role in Verlinde's discussion. 
The entropy $S$ takes different forms in different gravity theories. 
The first entropy-area relation of a black hole is given by the Bekenstein-Hawking formula $S=A/{4 l^2_p}$ \cite{f3}. 
However, this definition can be modified according to the result of the deformed HL gravity. 
The corrected entropy takes the form
\begin{equation}\label{eq5}
S=\frac{A}{4l^2_p}+\beta ln\frac{A}{l^2_p}
\end{equation}
Where $l^2_p$ is Planck area, $A$ is area of the holographic screen and $\beta$ is dimensionless constants of order unity.
Furthermore, the logarithmic term has also arisen from loop quantum gravity and the string theory etc. \cite{f47}\cite{f48}\cite{f49}.\\
\indent In the original paper \cite{f19}, the author suggested that the maximal storage space for information, or total number of bits, is proportional to the area $A$ where the holographic principle hold \cite{f46}, i.e.
\begin{equation}\label{eq6}
 N\sim A
\end{equation}
In this paper, we assume that on the horizon $N$ is proportional to the entropy $S$ rather than the area $A$ in this gravity theory as did in \cite{f23}, and study the consequences of this assumption. 
So with (\ref{eq5}), the total number of bits could be obtained
\begin{equation}\label{eq8}
 N=\frac{A}{4l^2_p}+\beta \ln \frac{A}{l^2_p}
\end{equation}
Equation (\ref{eq8}) is our newly proposed relation and would be used to derive the modified Friedmann equations of FRW universe 
together with the Debye model in next section.\\
\indent On the other hand, following Verlinde's scenario, we know that the gravity can be explained as an entropic force, 
which means that the gravity may have a statistical thermodynamics explanation. 
It should be noticed that the free particle's equipartition law of energy as one of assumptions was used to 
derive the Newton¡¯s law of gravitation. However, the equipartition law of energy does not hold at very low temperatures, 
and the equipartition law of energy derived from Debye model is in good agreement with experimental results.
Since the equipartition law of energy plays an important role in the derivation of entropic force, 
it should be modified for the very weak gravitational fields case which corresponds to very low temperatures. 
In \cite{f32}, Gao used the three dimensional Debye model to modify the entropic force.
Such modification can interpret the current acceleration of the Universe and avoid introducing any kind of dark energy.\\
\indent Here, we briefly mention the key points of MEF model \cite{f32}. One can modify the equipartition law of energy
\begin{equation}\label{eq9}
E=\frac{1}{2}\ NTD(x)
\end{equation}
where the one dimensional Debye function is defined as
\begin{equation}\label{eq10}
D(x)=\frac{3}{x^3}\int^x_0 \frac{y^3}{e^y-1}\mathrm{d}y
\end{equation}
and $x$ is related to the temperature $T$ as
\begin{equation}\label{eq11}
x=\frac{T_D}{T}=\frac{g_D}{g}
\end{equation}
It shows that when in strong gravitational fields, we have $x \ll 1$ and $D(x)\rightarrow 1$. In other words, 
it implies that the effect of Debye model would be so weak in the early universe. 
\indent Using the derivation method in \cite{f21}, one can find that the modified Raychaudhuri equation
is given by \cite{f32}
\begin{equation}\label{eq12}
4\pi (\rho+p) =-\left (\dot{H}-\frac{K}{a^2} \right ) \left [-2D(x)+\frac{3x}{e^x-1} \right ]
\end{equation}
Here $p$ is total pressure of cosmic fluids.
The energy conservation equation still holds in the MEF model as well, namely
\begin{equation}\label{eq13}
\dot{\rho}+ 3H(\rho+p)=0
\end{equation}
From Eqs.(\ref{eq12})and(\ref{eq13}), the corresponding Friedmann equation can be obtained. 
It is anticipated that Friedmann equation is also modified due to the correction term $-2D(x)+\frac{3x}{e^x-1}$ in Eq.(\ref{eq12}). If we consider a spatially flat universe ($K=0$), we have
\begin{equation}\label{eq14}
\frac{8\pi\rho}{3}=\int \left [-2D(x)+ \frac{3x}{e^x-1} \right ]\mathrm{d}H^2
\end{equation}
This MEF model is in fact a modified gravity model, which can describe the accelerating universe without dark energy \cite{f32}. Next, we will go further with this model and devote our efforts to obtaining the Friedmann equation in the deformed 
HL gravity by two methods.

\section{FRIEDMANN EQUATION FROM THE HAWKING TEMPERATURE}\label{SecC} 
\indent In this section, we reproduce the Friedmann equation from the Hawking temperature, introducing models of Debye and HL gravity. The homogeneous and isotropic universe model is described by the metric
\begin{equation}\label{eq15}
\mathrm{d}s^2=-\mathrm{d}t^2+a^2(t) \left (\frac{\mathrm{d}r^2}{1-kr^2}+r^2\mathrm{d}\Omega^2 \right )
\end{equation}
Here $\mathrm{d}\Omega^2$ is the line element of a two-dimensional unit sphere. Adopting a new coordinate $R=a(t)r$, the metric (\ref{eq15}) becomes
\begin{equation}\label{eq16}
\mathrm{d}s^2=h_{ab}\mathrm{d}x^a \mathrm{d}x^b+R^2\mathrm{d}\Omega^2
\end{equation}
With $x^0=t,x^1=r,h_{ab}=diag(-1,a^2/(1-kr^2))$ , the dynamical apparent horizon $R_A$ is determined by $h^{ab}\partial_aR\partial_bR=0$ , and is given by
\begin{equation}\label{eq17}
R_A=\frac{1}{\sqrt{H^2+k/a^2}}
\end{equation}
The temperature corresponding to the apparent horizon (Hawking temperature) is
\begin{equation}\label{eq18}
T=\frac{1}{2\pi R_A}
\end{equation}
Differentiate Eqs. (\ref{eq8}), (\ref{eq10}), (\ref{eq17}) and (\ref{eq18}), we have
\begin{equation}\label{eq19}
\mathrm{d}N=\left (\frac{1}{4l^2_p}+\frac{\beta}{Al^2_p}\right )8\pi R_A\mathrm{d}R_A
\end{equation}
\begin{equation}\label{eq20}
\mathrm{d}D(x)=\frac{1}{x}\left [-3D(x)+\frac{3x}{e^x-1}\right ]\mathrm{d}x
\end{equation}
\begin{equation}\label{eq21}
\dot{R}_A=-HR_A^3 \left (\dot{H}-\frac{K}{a^2} \right )
\end{equation}
\begin{equation}\label{eq22}
\mathrm{d}T=-\frac{\mathrm{d}R_A}{2\pi R_A^2}
\end{equation}
The change of the energy $\mathrm{d}E$ can be obtained from the equipartition law $E=\frac{1}{2}NTD(x)$.
Using Eqs. (\ref{eq19}), (\ref{eq20}), (\ref{eq21}) and (\ref{eq22}), we have (put $K=0$)
\begin{eqnarray}\label{eq23}
\mathrm{d}E&=&\frac{1}{2}ND(x)\mathrm{d}T+\frac{1}{2}TD(x)\mathrm{d}N+\frac{1}{2}TN\mathrm{d}D(x),
\\\nonumber
 &=& \left \{\left [\frac{2 \beta \left (1-2l^2_p\ln{\frac{A}{l^2_p}}\right )}{Al^2_p}-\frac{1}{2l^2_p}\right ]
 D(x)+\left (\frac{1}{4l^2_p}+\frac{\beta}{A}\ln{\frac{A}{l^2_p}}\right )\frac{3x}{e^x-1}\right \}
 \mathrm{d}R_A, \\\nonumber
 &=& \left\{\left [\frac{2 \beta (1-2l^2_p\ln{\frac{A}{l^2_p}})}{Al^2_p}-\frac{1}{2l^2_p}\right ]D(x)+
 \left (\frac{1}{4l^2_p}+\frac{\beta}{A}\ln{\frac{A}{l^2_p}}\right )\frac{3x}{e^x-1}\right\}
 (-HR^3_A\dot{H}\mathrm{d}t),
\end{eqnarray}
\indent Suppose that the energy-momentum tensor $T_{\mu\nu} $ of the matter in the universe has the form of a perfect fluid $T_{\mu\nu}=(\rho+p)u_\mu u_\nu+pg_{\mu \nu}$, where $u^\mu$ denotes the four-velocity of the fluid  and $\rho$ and $p$ are the energy density and pressure, respectively. The energy conservation law $\nabla_\mu T^{\mu \nu}=0$ gives the continuity equation in the form
\begin{equation}\label{eq24}
\dot{\rho}+3H(\rho+p)=0
\end{equation}
Following \cite{f15}, the amount of energy crossing the apparent horizon during the time interval $\mathrm{d}t$ is obtained
\begin{equation}\label{eq25}
\delta E=A(\rho+p)HR_A\mathrm{d}t
\end{equation}
Comparing (\ref{eq23}) and (\ref{eq25}), we reach
\begin{equation}\label{eq26}
\left\{ \left [ \frac{2\beta \left (1-2l^2_p \ln {\frac{A}{l^2_p}} \right )}{Al^2_p}-\frac{1}{2l^2_p} \right ]D(x)+
\left(\frac{1}{4l^2_p}+\frac{\beta}{A}\ln{\frac{A}{l^2_p}}  \right )\frac{3x}{e^x-1} \right \}
(-HR^3_A\dot{H}\mathrm{d}t)=4\pi R^2_A(\rho+p)HR_A\mathrm{d}t
\end{equation}
With the use of continuity equation (\ref{eq24}), we get
\begin{equation}\label{eq27}
\left\{ \left [ \frac{2\beta \left (1-2l^2_p \ln {\frac{A}{l^2_p}} \right )}{Al^2_p}-\frac{1}{2l^2_p} \right ]D(x)+
\left(\frac{1}{4l^2_p}+\frac{\beta}{A}\ln{\frac{A}{l^2_p}}  \right )\frac{3x}{e^x-1} \right \}
(\dot{H}H)=\frac{4\pi\dot{\rho}}{3}
\end{equation}
Integrating (\ref{eq27}), we derive the following equation
\begin{equation}\label{eq28}
\int \left\{ \left [ \frac{2\beta \left (1-2l^2_p \ln {\frac{A}{l^2_p}} \right )}{Al^2_p}-\frac{1}{2l^2_p} \right ]
D(x)+\left(\frac{1}{4l^2_p}+\frac{\beta}{A}\ln{\frac{A}{l^2_p}}  \right )\frac{3x}{e^x-1} \right \}
\mathrm{d}H^2=\frac{8\pi\rho}{3}
\end{equation}
In nature units, $l^2_p=1$. In the discussion from \cite{f24}, the effect of deformed HL gravity is appreciably reduced at small $\rho$ 
approximation. It has been shown that their modified Friedmann equation will go back to the standard Friedmann equation. 
Similarly, Eq.(\ref{eq28}) will degenerate to Gao's modified Friedmann equation \cite{f32} as long as $\rho$ is small enough, 
keeping all features in MEF model. In other words, with high energy density $\rho$ 
in early universe the effect of HL gravity model would be much stronger than the effect of MEF model. 
Contrarily, with low energy density $\rho$ in late universe the latter one would be the dominant factor. 
So it has been shown that Eq.(\ref{eq28}) is better than other models in describing the evolution of the universe.

\section{FRIEDMANN EQUATION FROM THE UNRUH TEMPERATURE}\label{SecD}
\indent By use of the holographic principle together with the equipartition law of energy and the Unruh temperature, 
Cai derived the Friedmann equation of a Friedmann-Robertson-Walker universe \cite{f22}. In this part, motivated by Cai's method, another modified 
Friedmann equation with the idea of Debye model and the modified HL gravity is obtained.\\
\indent In Verlinder's assumption, an observer in an accelerated frame has the Unruh temperature \cite{f43}
\begin{equation}\label{eq29}
T=\frac{a_r}{2\pi}
\end{equation}
the acceleration for a radial comoving observer at r, namely at the place of the screen, is
\begin{equation}\label{eq30}
a_r=-\frac{\mathrm{d}^2R_A}{\mathrm{d}t^2}=-\ddot{a}r
\end{equation}
where the negative sign arises because we consider the acceleration is caused by the matter in the spatial region enclosed
by the boundary $\partial V$. A simple calculation leads to
\begin{equation}\label{eq31}
T=-\frac{\ddot{a}r}{2\pi}
\end{equation}
This is different from the Hawking temperature case. For the sake of clarity, we list the relative quantities
\begin{equation}\label{eq32}
N=\frac{A}{4l^2_p}+\beta \ln \frac{A}{l^2_p}
\end{equation}
\begin{equation}\label{eq33}
E=\frac{1}{2}\ NTD(x)
\end{equation}
Comparing(\ref{eq31}),(\ref{eq32}) and (\ref{eq33}), we reach
\begin{equation}\label{eq34}
E=-\frac{\ddot{a}r}{4\pi}\ (\frac{A}{4l^2_p}+\beta \ln \frac{A}{l^2_p})\ D(x)
\end{equation}
On the other hand, using Eqs.(\ref{eq15}) and (\ref{eq25}), the active gravitational mass is defined as \cite{f22}
\begin{equation}\label{eq35}
M=2\int_v(T_{\mu\nu}- \frac{1}{2} Tg_{\mu\nu})u^\mu u^\nu\mathrm{d}V=\frac{4\pi a^3 r^3}{3} (\rho+3p)
\end{equation}
Because $E=M$, we have
\begin{equation}\label{eq36}
-\frac{\ddot{a}r}{4\pi}\ \left (\frac{A}{4l^2_p}+\beta \ln \frac{A}{l^2_p}\right )\ D(x)=\frac{4\pi a^3 r^3}{3} (\rho+3p)
\end{equation}
or
\begin{equation}\label{eq37}
-\frac{1}{4\pi R_A^2}\left (\frac{A}{4l^2_p}+\beta \ln \frac{A}{l^2_p} \right )\frac{\ddot{a}}{a} D(x)=\frac{4\pi}{3} (\rho+3p)
\end{equation}
Note that we have used
\begin{equation}\label{eq38}
R_A=ar
\end{equation}
in above derivation.  Then we derive
\begin{equation}\label{eq39}
-\left (\frac{1}{4l^2_p}+\frac{\beta}{A} \ln \frac{A}{l^2_p} \right )\frac{\ddot{a}}{a} D(x)=\frac{4\pi}{3} (\rho+3p)
\end{equation}
This is just the modified acceleration equation for the dynamical evolution of the FRW universe. Multiplying $\dot{a}a$ on both sides of Eq.
(\ref{eq39}), and using the continuity equation(\ref{eq24}), we integrate the resulting equation and obtain (K=0)
\begin{equation}\label{eq40}
H^2=\left (\frac{\dot{a}}{a} \right )^2\left [\frac{D(x)}{4l^2_p}+\frac{\beta D(x)}{A \rho}\left (\ln \frac{A}{l^2_p} \right )\int\frac{\mathrm{d}(\rho a)^2}
{a^2} \right ]=\frac{8\pi\rho}{3}
\end{equation}
In nature units, $l^2_p=1$. In this way we derive another modified Friedmann equation of FRW universe by considering HL gravity and Debye model from entropic force.
In absence of the correction terms$(\beta=0)$, one recovers the approximate MEF model. Since the 
logarithmic term can be ignored only when $a$ is very small, the corrections make sense only at early stage of the universe where
$a \rightarrow 0$. When the universe becomes large ($a \gg 1$), Eq.(\ref{eq40}) would reduce to the approximate Gao's modified Friedmann equation. 
Since the effect of this Debye model could be obvious enough, 
it can sufficiently describe the accelerating universe without the assumption of dark energy \cite{f32}. 
Although Eq.(\ref{eq40}) is different from Eq.(\ref{eq28}),  it has the same conclusion as the previous discussion.

\section{conclusion}\label{SecE}
\indent In summary, Ho\v{r}ava-Lifshitz (HL) theory is regarded as a UV complete candidate for general relativity, 
while MEF model can interpret the current acceleration of the universe without invoking any kind of dark energy. 
Considering the results from HL theory, 
a more precise relation between the number of bits and area in the holographic system is proposed. 
With this corrected relation and Debye model, 
two modified Friedmann equations are derived 
by using two different methods in entropic force. 
These modified Friedmann equations are derivated out from the Hawking temperature and the Unruh temperature separately. 
The Friedmann equation in the deformed HL gravity and Einstein gravity are very different in early universe, 
but they give the same result in late universe. 
On the other hand, the modified Friedmann equation considering Debye model and Einstein gravity vary in late universe, but agree with each other in early universe. 
Our discussion gives different modified Friedmann equations but both of them would degenerate to MEF model under certain conditions. 
These new equations could be better in describing the whole evolution of the universe. 
The results may be useful to further investigations of the holographic properties and modified Friedmann equations for different gravity theories.

\section{ackonwledgement}
The authors would like to thank Dr. Changjun Gao for his helpful discussions.

\end{document}